\documentclass[review]{elsarticle}
\graphicspath{{img/}}
\usepackage[margin=1.5in]{geometry}
\usepackage{amsthm,amssymb,amsmath}

\usepackage[english]{babel}
\usepackage[utf8]{inputenc}
\usepackage{amsmath}
\usepackage{graphicx}
\usepackage[colorinlistoftodos]{todonotes}
\usepackage{color}
\usepackage{algorithmic}
\usepackage{color,soul}
\usepackage{algorithmic}
\usepackage{comment}
\usepackage{natbib}

\usepackage{booktabs}

\usepackage{algorithm}
\usepackage{subcaption}
\usepackage{caption}
{
	{
		\theoremstyle{plain}
		
	}
	{
	\theoremstyle{plain}
	
}		
	{
		\theoremstyle{plain}
		
	}
	{
		\theoremstyle{plain}
		
	}
{
		\theoremstyle{remark}
		
	}

{
		\theoremstyle{plain}
		
	}}

\usepackage{url}            

\usepackage{algorithm}
\usepackage{algorithmic}

\usepackage{algorithmic}
\theoremstyle{theorem}

\theoremstyle{remark}

\begin{document}

\begin{frontmatter}

\title{ Prescribing Optimal Health-Aware Operation for Urban Air Mobility with Deep Reinforcement Learning}

\author[rvt1,rvt2]{Mina Montazeri}
\author[rvt]{Chetan S. Kulkarni}
\author[rvt1]{Olga Fink\corref{cor1}}
\cortext[cor1]{Corresponding author, email address:olga.fink@epfl.ch.}
\address[rvt1]{Laboratory of Intelligent Maintenance and Operation Systems, EPFL, Switzerland.}
\address[rvt2]{  Urban Energy Systems Laboratory, Swiss Federal Laboratories for Materials Science and Technology, Dübendorf, Switzerland.}
\address[rvt]{KBR, Inc., NASA Ames Research Center, Mountain View, CA 94035, USA}

\begin{abstract}
Urban Air Mobility (UAM) aims to expand existing transportation networks in metropolitan areas by offering short flights either to transport passengers or cargo. Electric vertical takeoff and landing aircraft powered by lithium-ion battery packs are considered promising for such applications. Efficient mission planning is crucial, maximizing the number of flights per battery charge while ensuring completion even under unforeseen events. As batteries degrade, precise mission planning becomes challenging due to uncertainties in the end-of-discharge prediction. This often leads to adding safety margins, reducing the number or duration of potential flights on one battery charge. While predicting the end of discharge can support decision-making, it remains insufficient in case of unforeseen events, such as adverse weather conditions. This necessitates health-aware real-time control to address any unexpected events and extend the time until the end of charge while taking the current degradation state into account. 
This paper addresses the joint problem of mission planning and health-aware real-time control of operational parameters to prescriptively control the duration of one discharge cycle of the battery pack. We propose an algorithm that proactively prescribes operational parameters to extend the discharge cycle based on the battery's current health status while optimizing the mission. The proposed deep reinforcement learning algorithm facilitates operational parameter optimization and path planning while accounting for the degradation state, even in the presence of uncertainties. Evaluation of simulated flights of a  National Aeronautics and Space Administration (NASA) conceptual multirotor aircraft model, collected from Hardware-in-the-loop experiments, demonstrates the algorithm's near-optimal performance across various operational scenarios, allowing adaptation to changed environmental conditions. The proposed health-aware prescriptive algorithm enables a more flexible and efficient operation not only in single aircraft but also in fleet operations,  increasing the overall system throughput. 
\end{abstract}
\end{frontmatter}

\section{Introduction}
\label{sec:Introduction}
Urban Air Mobility (UAM) represents a paradigm shift in transportation, aiming to revolutionize urban mobility in densely populated metropolitan areas. With the goal of addressing challenges such as traffic congestion, limited accessibility, and environmental concerns, UAM offers a promising solution through short-distance flights for both passengers and cargo~\cite{vascik2018analysis,sripad2021promise}. Among the diverse range of aircraft types being explored for UAM applications, electric vertical takeoff, and landing (eVTOL) aircraft, propelled by advanced lithium-ion battery packs, have emerged as a particularly promising option \cite{ZHAO2024107386,liu2023flying}. These cutting-edge aircraft possess remarkable potential in terms of their ability to provide highly efficient, environmentally friendly, and exceptionally quiet air transportation within urban environments \citep{saipradeepenergy1}.

In the context of UAM, efficient mission planning plays a crucial role in optimizing the utilization of eVTOL aircraft and ensuring their viability as a mode of transportation. The primary objective is to plan missions to maximize the number of flights that can be performed on a single battery charge while ensuring the successful completion of each flight or mission, even in the face of unforeseen events or disruptions such as adverse weather conditions \cite{hill2020uam,guerreiro2019mission}. 
Various methods have been explored for path planning, including guidance algorithms using Markov Decision Processes (MDP) \cite{bertram2019online,chen2023real}, 
Reinforcement Learning (RL) \cite{xue2023dynamic,bernini2024reinforcement,lee2023real}, Particle Swarm Optimization (PSO) \cite{HUANG2023105942,jia2022double,li2024multi}, visibility graph algorithms \cite{semanjski2022aircraft}, Dynamically Directed Graph Algorithm (DDGA) \cite{ZHAO2024107386}, evolutionary algorithm \cite{zhang2024novel}, and chimp optimization
algorithm \cite{chen2023uav}. 
However, a common limitation of these approaches is the neglect of important factors such as battery aging and the absence of joint mission planning with real-time control of operational parameters, which are critical in real-world UAM systems~\cite{salinas2023battery}. As batteries degrade over time, accurately predicting the end-of-discharge becomes increasingly challenging, leading to increased uncertainty in estimating the available energy for flight operations \cite{biggio2022dynaformer,mazzi2024lithium}. These uncertainties pose significant challenges to mission planning, necessitating innovative approaches to effectively address them.

 While previous research has investigated the ability to predict the end-of-discharge of batteries using both model-based \cite{meng2018overview,bartlett2015electrochemical} and data-driven approaches \cite{jiao2021more,aykol2020machine,biggio2022dynaformer}, the efficiency of this prediction in determining optimal aircraft trajectories remains unexplored. Finding optimal trajectories becomes more complex in the face of unforeseen events or disturbances, such as adverse weather conditions, where the decision algorithm must discern whether its low performance is due to the current battery health state or weather conditions. As such, there is a pressing need to develop advanced algorithms and methodologies that can effectively address the joint challenges of mission planning and health-aware real-time control of operational parameters in the context of battery aging and adverse weather conditions.

This paper proposes a deep reinforcement learning algorithm designed to address the intricate challenge of simultaneous mission planning and health-aware real-time control of operational parameters for eVTOL aircraft. The algorithm takes into account crucial factors such as battery aging and unforeseen events. Effective decision-making relies on a comprehensive understanding of how battery aging and unexpected events impact discharge trajectories, with variations specific to each mission. The proposed algorithm leverages the recently introduced Dynaformer algorithm, a transformer-based architecture \cite{biggio2022dynaformer}, to forecast the voltage profile of the entire flight. This prediction relies on a brief observation time period (20s) during which both the current and voltage profiles are collected. Following this observation phase, the algorithm implicitly infers battery aging information through the encoder part of the Dynaformer and conditions the prediction of the discharge trajectory based on the currently planned flight profile. The decision-making process leverages deep Q-Learning, a form of deep reinforcement learning that enables the aircraft to make optimal decisions, in scenarios where the action space is continuous. Specifically, the algorithm empowers the aircraft to make strategic decisions, such as selecting destinations and their order, as well as tactical decisions regarding the altitude for each destination, all while considering the information on the health condition.

To comprehensively evaluate the efficiency of our proposed algorithm, we conducted extensive simulations using a  National Aeronautics and Space Administration (NASA) conceptual multirotor aircraft model. These simulations are executed within a hardware-in-the-loop experimental framework, encompassing realistic flight scenarios and diverse environmental conditions, including various wind conditions~\cite{sai2020wind, gano, kulkarni2022}. Our assessment covered a range of mission scenarios to gauge the performance of our proposed health-aware prescriptive algorithm. These mission scenarios represent three distinct flight scenarios. The first mission scenario comprises a single flight, while the second scenario involves the aircraft reaching multiple destinations, each assigned with varying levels of priority. The objective of this mission is to reach as many high-priority destinations as possible. In the third scenario, the mission requires the aircraft to reach all predefined destinations with the minimum number of charging cycles. 

Our research outcomes have significant implications for the operation of eVTOL aircraft within UAM systems. Our proposed algorithm not only enhances the operation of individual aircraft for diverse mission scenarios but also contributes to the overall efficiency and sustainability of fleet operations. 

In each of the three scenarios, the algorithm successfully achieved its objective by determining optimal paths and altitudes for each flight, taking into account the prevailing battery health status, wind conditions, and scenario priorities.
Therefore, optimizing operational parameters and path planning, while considering battery degradation and uncertainties, is crucial for ensuring the long-term viability and success of UAM as a reliable and scalable transportation solution.

\section{Methodology}
\label{Sec:Meth}
\subsection{Methodological Framework}
The goal of this work is to develop a decision-making model capable of simultaneously handling mission planning and health-aware real-time control of operational parameters in various mission scenarios.
The key innovation lies in the incorporation of the battery's health status and the ability to respond to unforeseen events, such as adverse weather conditions, within our decision-making algorithm.
In this case, to make accurate decisions, it is crucial to understand how the battery state and unexpected events affect the discharge curve in a specific mission.
To accomplish this, we employ the health-aware discharge prediction module, which consists of a Dynaformer to forecast the voltage profile of the entire flight. This forecasting relies on a brief observation time period (20s) during which both the current and voltage profiles are collected, as detailed in~\cite{biggio2022dynaformer}. 
Subsequently, leveraging the discharge prediction, we introduce a decision-making agent employing deep reinforcement learning (DRL). This empowers the aircraft controller to acquire optimal decisions even in the absence of precise knowledge about flight circumstances, such as adverse weather conditions. Consequently, the aircraft can determine destinations (mission planning), their sequence (mission planning), and the altitude for each destination (control of operational parameters) using DRL while considering the current system health state for decision-making. To ensure optimal decision-making, it is crucial to appropriately define the state and reward.  In this work, the state vector encompasses key information, including the predicted voltage discharge curve, implicitly inferred battery health representation, next destination locations, and the count of destinations reached. This comprehensive state representation provides the necessary environmental information for making optimal decisions. Regarding the reward, the aircraft receives positive reinforcement upon successfully reaching each destination. However, substantial penalties are imposed if the battery voltage falls below the end-of-discharge level at any point during the mission, before reaching the destination, or if the aircraft arrives at the charging station. This penalty structure is designed to incentivize the aircraft to efficiently reach multiple destinations within a single charging cycle, provided the battery voltage remains above the end-of-discharge threshold.

To assess the effectiveness of our proposed algorithm, we conducted simulations employing a NASA conceptual multirotor aircraft utilizing the Li-ion battery pack~\cite{sai2020wind, gano, kulkarni2022}. These simulations were executed within a hardware-in-the-loop experimental framework, encompassing realistic flight scenarios with various battery health states and varying operational wind conditions. Since the health-aware discharge prediction module is available at the level of a single cell, the simulator data is then scaled down to a single 18650 cell. All subsequent calculations are performed using the scaled values. 

\begin{figure}[h!]
	\centering
\includegraphics[width=1\linewidth]{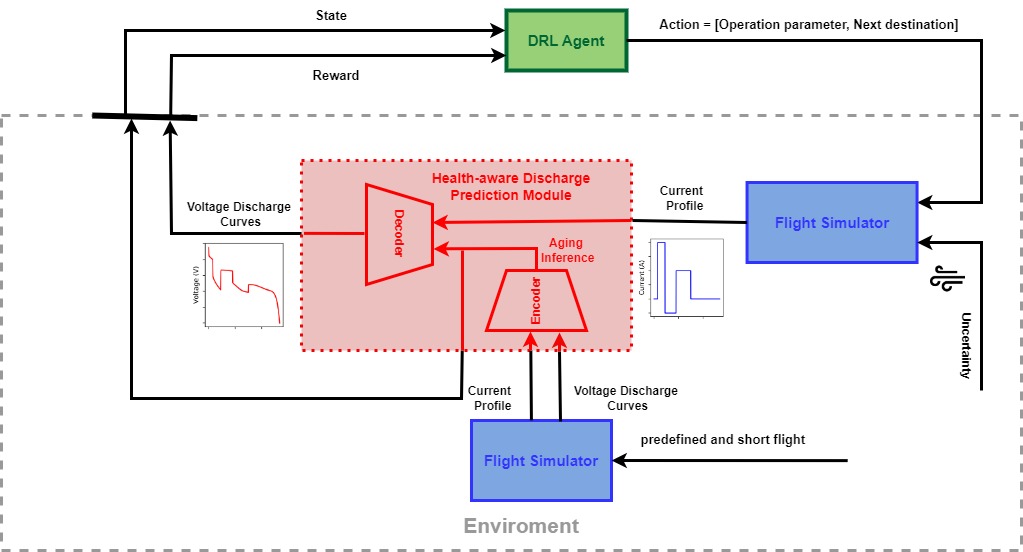}
	\caption{   Overview of the proposed health-aware optimal operation method for urban air mobility. 
 }
	\label{Fig:RL}
\end{figure}

An overview of the proposed health-aware optimal operation framework is illustrated in Figure~\ref{Fig:RL}. In the following, the main building blocks of this figure are explained in detail.
\subsection{Health-Aware Discharge Prediction Module}

To make an optimal decision for joint mission planning and real-time control of operation parameters, we first need to predict the voltage discharge curve of the battery under the current battery health conditions. To accomplish this, we develop the health-aware discharge prediction module (red module in Figure \ref{Fig:RL}) that predicts the voltage discharge curve by considering the impact of battery health on the voltage discharge trajectories. To design this module, we employ a Dynaformer that forecasts the voltage profile of the entire flight, relying on a brief observation time period (20s) during which both the current and voltage profiles are collected \cite{biggio2022dynaformer}.

In the first step, a substantial and diverse dataset is generated, incorporating synthetic voltage curves derived from distinct input current profiles and various health conditions, covering a broad spectrum of combinations of health parameters. This involves the utilization of the recently introduced open-source NASA simulator of the electrochemical battery model, implementing the single-cell battery model outlined in \cite{Daigle2013Batts,daigle2016end}. The foundational element required for constructing the dataset is a set of parameters representing the battery's degradation level. This study concentrates on two such parameters: $q_{max}$, capturing the total amount of available active Li-ion, and $R_0$, accounting for the increase in internal resistance.
To create this comprehensive dataset, a significant number of $(q_{max},R_0)$ pairs and current profiles are sampled, ensuring that the resulting voltage curves generated by the simulator display maximum diversity. For each flight
the values of $q_{max}$ and $R_0$ were randomly drawn from uniform distributions with supports between 5000 C and 8000 C and 0.017 $\Omega$ and 0.45 $\Omega$, respectively.
 All other parameters and initial conditions are maintained at typical values assumed for Li-ion 18650 batteries (see table 1 in \cite{Daigle2013Batts}).

Dynaformer employs a transformer neural network architecture, as depicted in Figure~\ref{Fig:RL}.  In line with observations from  \cite{biggio2022dynaformer}, the encoder component of Dynaformer implicitly learns the health state of the battery from observations, even in the absence of ground truth information regarding degradation conditions, which is typically unavailable in real-world applications.  Specifically, the encoder infers the health state with only the initial segment of current/voltage profiles. In this implementation, the first 20 seconds of both the current and voltage profiles are used as inputs to the encoder for health inference. Subsequently, based on the encoder's output, the decoder predicts the full voltage discharge trajectory extending until the end of discharge, conditioned on any given input current load profile.
Overall, the Dynaformer module plays a crucial role in our framework for health-aware discharge prediction, facilitating effective decision-making for the operational flight's parameters with a focus on health awareness.

\subsection{Deep Reinforcement Learning Agent}
\label{sec:Reinforcement learning}
\begin{figure}[h!]
	\centering
\includegraphics[width=0.85\linewidth]{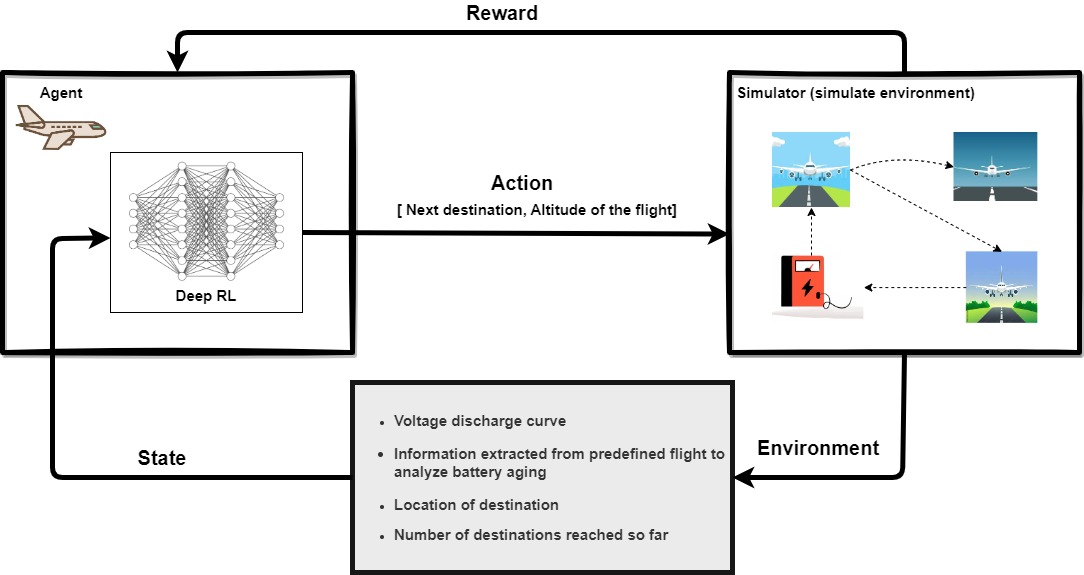}
	\caption{  The agent-environment interaction in deep reinforcement learning method
 }
	\label{Fig:RL4}
\end{figure}

The deep reinforcement learning agent 
(green module in Figure \ref{Fig:RL}) we developed focuses on optimizing critical operational decisions for aircraft, with a primary emphasis on the health state of the battery. In this study, we explicitly consider two key operational parameters: determining the optimal altitude and establishing the most efficient sequence for reaching destinations. While other variables like speed could be a potential additional parameter, our investigation specifically centers on these two decisions within the framework of the algorithm. Figure \ref{Fig:RL4} provides a comprehensive overview of agent-environment interaction in the DRL method.

To ensure the successful implementation of DRL, it is essential to define the learning environment. Within our framework, the environment comprises three key components: the health-aware discharge prediction module, the flight simulator, and the adverse weather conditions sampler. 
The state vector integrates information from the environment, encompassing the following data:
\begin{itemize}
\item The predicted voltage discharge curve for the mission up to the next destination, which is obtained by inputting the simulated current curve from the flight simulator into Dynaformer.
\item Implicitly inferred battery health representation, which is derived by inputting the initial segment of the current/voltage profiles from the flight, approximately  20 seconds in duration, into the encoder. The encoder's output provides the inferred health status.
\item The next destination locations.
\item The count of destinations reached so far.
\end{itemize}
Thus, the state vector can be expressed as follows:
\begin{equation}
s=[PC_1, PC_2 , V , (x,y) ,  i ],
\label{eq:state}
\end{equation}
where $PC_1$ and $PC_2$ represent the first two principal components of the Dynaformer's encoder embedding. These components have been demonstrated to exhibit strong correlations with degradation parameters $q_{max}$ and $R_0$, capturing the total amount of available active Li-ions and
the increase in internal resistance, respectively \cite{biggio2022dynaformer}. Here, $V$ denotes the voltage of the battery at the present destination. 
The aircraft's coordinates at the current destination are given by $(x,y)$. 
The variable $i$ represents the number of destinations that the aircraft has reached up to the current destination. 
The aircraft's action involves determining  the next destination and the altitude at which it will fly to reach the destination, which can be expressed  as follows:
\begin{equation}
a =[(x',y'),h ],
\label{eq:action}
\end{equation}
where $(x',y')$ represents the coordinates of the next destination, and $h$ denotes the altitude at which the aircraft must fly to reach that destination. In the algorithm proposed in this study,  the aircraft is rewarded upon successfully reaching each destination, with rewards based on pre-determined priority. However, if the battery voltage drops below the end-of-discharge level at any destination or before reaching the destination, the aircraft incurs a significant penalty. Additionally, the aircraft incurs a penalty upon reaching the charging station, albeit smaller compared to when the battery voltage falls below the end-of-discharge level. This penalty structure encourages the aircraft to reach multiple destinations within a single charging cycle,  maximizing both performance and efficiency. 
Thus, the reward can be formulated as follows:
\begin{equation}
r=\begin{cases}
    \alpha, & \text{if  Battery's voltage$>EOD$},\\
    -\beta, & \text{if aircraft reaches charging station $\&$ Battery's voltage$>EOD$},\\
    -\gamma, & \text{if Battery's voltage$<EOD$},
  \end{cases}
\label{eq:reward}
\end{equation}
where $\alpha$, $\beta$, and$\gamma$ are three positive parameters and $\gamma >>\beta$. The term $EOD$ denotes the end of discharge, defined as the time when a battery reaches a certain pre-specified voltage cut-off value, indicating full discharge.

Given the continuous nature of the state representation (where all elements of the state vector, including the principal components and the battery voltage in state space,  are continuous parameters as per (Equation \ref{eq:state})), and the discrete nature of the action space (as defined in Equation \ref{eq:action}), we employ Deep Q-Learning (DQL). DQL is a popular method for solving reinforcement learning problems \cite{mnih2015human} and serves as a variant of the Q-Learning algorithm, utilizing a deep neural network to approximate the Q-function. 
Q-Learning is a model-free algorithm that learns an optimal policy by iteratively improving the Q-values of state-action pairs. The Q-value for a state-action pair $(s,a)$ represents the expected cumulative reward achieved by taking action $a$ in state $s$ and then following the optimal policy. Q-Learning updates these Q-values using the Bellman equation:
\begin{align*}
Q(s,a) &\leftarrow Q(s,a) + \alpha [r + \gamma \max_{a'} Q(s',a') - Q(s,a)],
\end{align*}
where $s'$ is the next state, $r$ is the reward, $\alpha$ is the learning rate, and $\gamma$ is the discount factor.

Deep neural networks (DNN) serve as powerful function approximators capable of learning complex mappings between inputs and outputs. By combining Q-Learning with DNN, Deep Q-Learning can effectively solve reinforcement learning problems characterized by high-dimensional state spaces and discrete action spaces.
The Q-values are computed using a deep neural network with state $s$ as the input and action $a$ as the output:
\begin{align*}
Q(s,a;\theta) &= f_\theta(s,a)
\end{align*}
where $f_{\theta}$ is the neural network parameterized by $\theta$. 
The loss function corresponds to the mean squared error between the predicted Q-values and the target Q-values:
\begin{align*}
\operatorname{loss} &= \mathbb{E}[(y_t - Q(s_t,a_t;\theta))^2]
\end{align*}

To stabilize training, the target Q-values $y_t$ are computed using a separate target network parameterized by $\theta^-$:
\begin{align*}
y_t &= r_t + \gamma \max_{a'} Q(s_{t+1},a';\theta^-)
\end{align*}
The target network parameters $\theta^-$ are periodically copied from the Q-network weights $\theta$. This periodic copying ensures that the target Q-values remain fixed for a  certain duration, enabling the Q-network to adapt slowly to changing targets.
The Deep Q-Learning algorithm uses an $\epsilon$-greedy exploration strategy to balance exploration and exploitation. 
In essence, the algorithm aims to learn an optimal policy by iteratively improving the Q-values through the integration of experience replay, target networks, and DNN.  

\subsection{Flight Simulator}
\label{sec:Simulator}
To generate distinct current profiles under various battery health conditions and wind conditions as the input of the health-aware discharge prediction module, we simulated flights using a NASA conceptual multirotor aircraft model obtained from hardware-in-the-loop experiments \cite{kulkarni2022} 
(blue module in Figure \ref{Fig:RL}). We employ the electro-chemistry-based model of a Li-ion battery pack, developed by Daigle and Kulkarni \cite{Daigle2013Batts}. This model, along with associated prognostics algorithms has undergone rigorous verification and validation procedures, including previous research conducted on electric unmanned aerial vehicles \cite{HoggeIJPHM18}. We specifically focus on Li-ion 18650 batteries with an average nominal voltage of 3.7V and a nominal capacity of 2200mAh. A pack size of 10S50P, as discussed in \cite{kulkarni2022}, is used as a reference battery pack. However, the model is general enough that, with some modifications, it may be applied to different battery chemistries \cite{kulkarni2022}.

In the hardware-in-the-loop experiments, cells, and packs are subjected to charge/ discharge profiles expected during flights. The MACCOR Battery tester system is programmed to run a  power demand profile similar to the simulated trajectory for analyzing battery performance at different operating conditions for comprehensive analysis \cite{kulkarni2022}. The Series 4000 MACCOR Battery Tester system comprises essential components, including a test cabinet, a computer, and software for both the tester and data analysis. Inside the test cabinet,  microprocessor controllers efficiently manage tests for dynamic loading of the battery packs and collect data. Each controller board can support multiple test channels, from 5V,20A upto 100V, 250A. Individually controllable Programmable loads and power supplies are integrated into the test cabinet. This design allows for the independent operation of each test channel, enabling simultaneous testing of various parameters. Test programming is performed on the computer through a user-friendly Graphical User Interface (GUI) \cite{kulkarni2022}.

The dataset used for training the models is the High-Intensity Radiated Fields (HIRF)  Data $\#$ 15 ~\cite{HIRF2015}. The dataset utilized in this study comprises experimental data from HIRF tests conducted on the Edge 540, a small electric unmanned aerial fixed-wing vehicle (e-UAV). The e-UAV employed in this research represents a scaled-down version, measuring 33\% of the size, of the Zivko Aeronautics Inc. Edge 540-T tandem seat aerobatic aircraft. To perform ground-based testing of the Edge 540-T hardware and software, the vehicle was securely fastened in the HIRF test chamber. For further details regarding the HIRF Chamber, reference can be made to a previously published report on UAS radio frequency emissions testing \cite{ely2011radiated}. The aircraft was positioned on expanded polystyrene blocks, meticulously centered within the chamber. The aircraft's powertrain, equipped with a propeller, was operated while the vehicle remained anchored to the chamber wall using a steel cable. The motor and actuators of the aircraft were controlled remotely from a separate room, utilizing the same radio control system employed during flight tests. Data collection was performed during the experiments with the aircraft operating in manual/auto control mode. {{Operating flight profiles, though similar, exhibited variations based on factors such as wind conditions,  and temperature. 
\section{Case Study}
In this section, we present the results of the proposed health-aware optimal operation method designed for urban air mobility outlined in Section \ref{Sec:Meth}. Our evaluation focuses on conceptual multirotor aircraft model simulated flights obtained from hardware-in-the-loop experiments as detailed earlier. 
To assess the performance of the proposed framework under a range of realistic flight conditions, we conduct simulations for various mission scenarios at various altitudes, including 500 m (1640 ft), 1000 m (3280 ft), 2000 m (6561 ft), and 3000 m (9842 ft). In addition, different wind speeds are considered, including headwinds of 13, 26, and 39 kts, as well as tailwinds of 13, 26, and 39 kts. Further details about the dataset can be found in Section \ref{sec:Simulator} and \ref{apB}.
In the following, the performance results of the proposed DRL agent across various mission scenarios are presented.
\subsection{Mission Scenarios}
\label{sec:Mission Scenarios}
We explore various mission scenarios to assess the performance of our proposed health-aware prescriptive DRL algorithm.  These scenarios encompass three distinct types, each representing a different flight mission. The first mission consists of a single flight, while the second and third missions involve multiple flights to various destinations. Each mission will be discussed in further detail below.
\subsubsection{Single Flight Mission}
In the first scenario, we consider a mission with a single flight, aiming to determine the optimal altitude as the operating parameter for the aircraft under different adverse wind conditions and with different health conditions. Optimizing the flight's altitude can significantly extend the time to the end of discharge for the battery, thereby allowing for longer flight times. Figure \ref{fig:altitude1} illustrates the effect of varying altitudes on the corresponding voltage discharge curve and current profile.
\begin{figure}[h!]
     \centering
     \begin{subfigure}[b]{0.45\textwidth}
         \centering
         \includegraphics[width=1\linewidth]{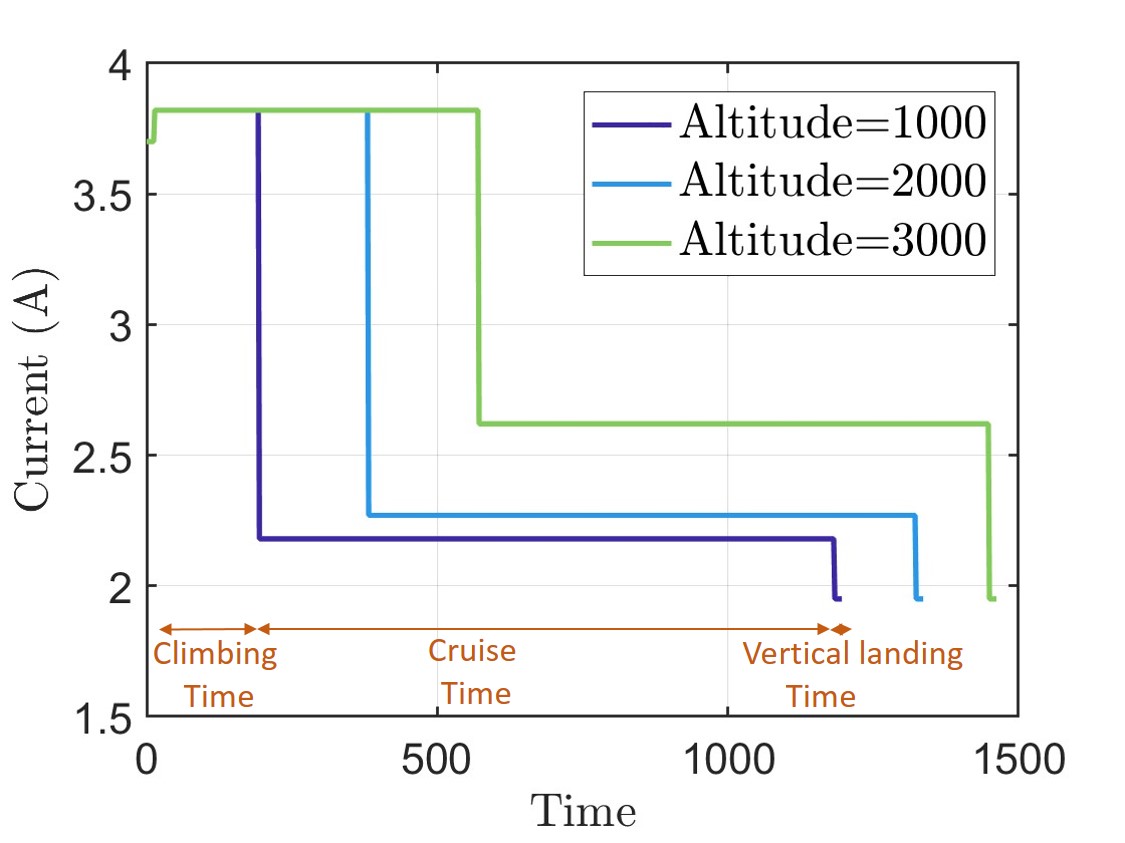}
         \caption{  current profile for different altitudes}
         \label{current1}
     \end{subfigure}
     \hfill
     \begin{subfigure}[b]{0.45\textwidth}
         \centering
         \includegraphics[width=1\linewidth]{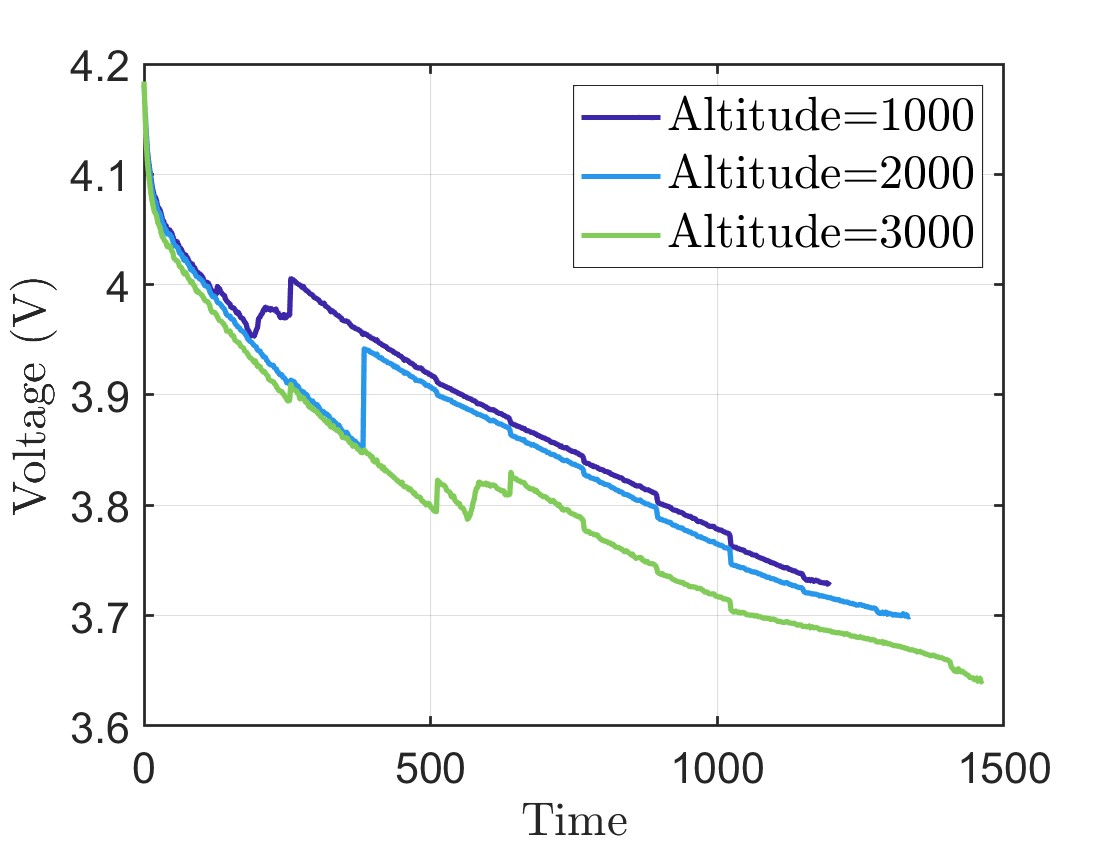}
         \caption{voltage discharge curves for different altitudes}
           \label{R1}
     \end{subfigure}
              \caption{ Effect of altitude on the voltage discharge curves and current profile for a specific flight. The orange lines in (a) show the climbing, cruise, and vertical landing times for a flight at an altitude of 1000 meters.}
        \label{fig:altitude1}
\end{figure}
Nevertheless, the end of discharge is influenced by additional factors, including adverse weather conditions and the degradation state of the battery. Adverse weather conditions may lead to increased energy consumption and diminished battery performance. Since weather conditions are typically unpredictable, they can lead to unfinished flights or prevent the aircraft from returning to the charging station. Similarly, the degradation state of the battery, reflecting the extent to which the battery has lost capacity over time, significantly impacts the end-of-discharge. Figures \ref{fig:degradation} and \ref{fig:wind1} illustrate the effects of degradation parameters (total available lithium ions i.e. $q_{max}$ and internal resistance i.e. $R_0$) and adverse wind conditions on the end of the discharge. 
To ensure an accurate prediction of the end of discharge for a single flight, it is crucial to consider both the current degradation state of the battery and the uncertainties arising from adverse weather conditions. To mitigate potential adverse events, such as unfinished flights or loss of the aircraft, proactive reactions to both health conditions and adverse events in real time are essential, as proposed in this algorithm.

In Figure \ref{fig:altitude}, we present the results of the proposed framework for two distinct single-flight missions: mission \#1 and mission \#2. The key difference between these missions is that the second mission is twice as long as the first. The altitude chosen in each learning episode by the aircraft is illustrated in Figure \ref{sim1}. This figure shows that the aircraft learns to optimize its altitude choice based on the mission duration. Specifically, in mission \#1, the optimal altitude is determined to be 500 meters, while in mission \#2, it increases to 1000 meters. This behavior can be attributed to the observation depicted in Figure \ref{current1}, where the climbing time increases with higher flight altitudes, while the cruise time decreases. Therefore, for shorter flights, flying at a lower altitude proves to be optimal, while for longer flights, the aircraft finds it advantageous to endure additional climbing time in exchange for reduced cruise time at higher altitudes.
\begin{figure}[h!]
     \centering
     \begin{subfigure}[b]{0.32\textwidth}
         \centering
         \includegraphics[width=1\linewidth]{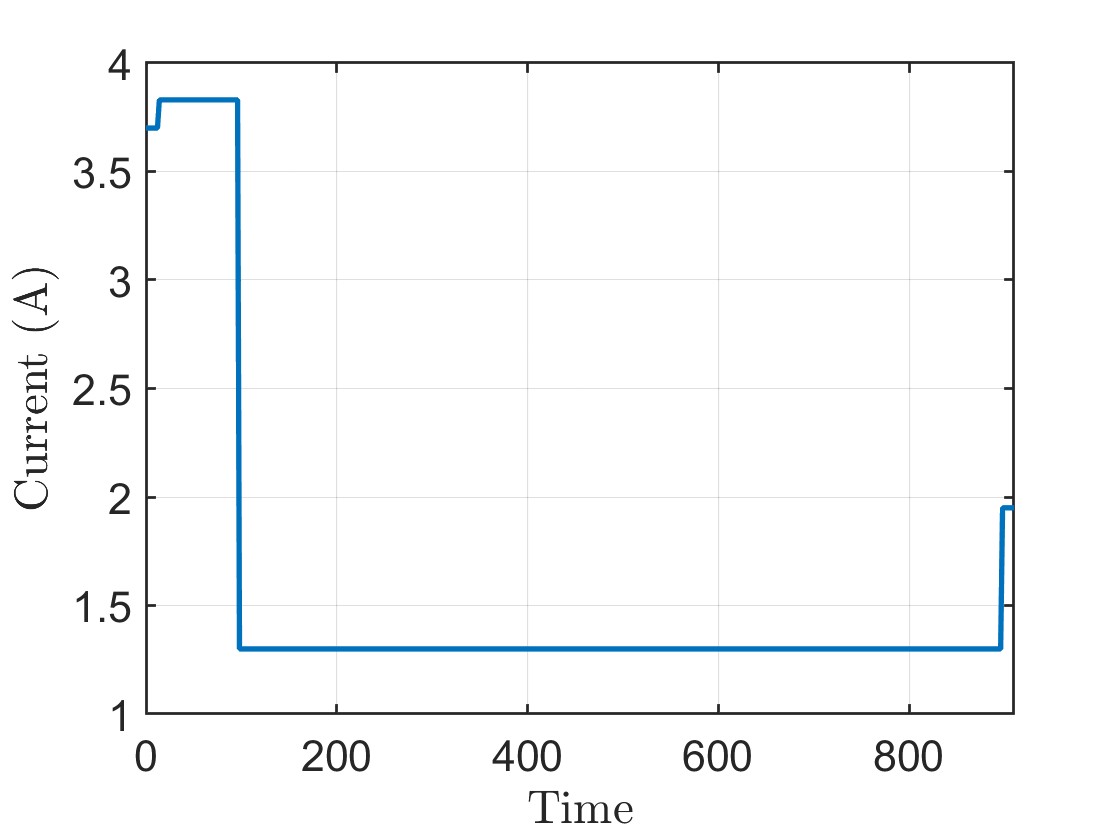}
         \caption{  Current profile for specific flight {\color{white}....}}
         \label{current}
     \end{subfigure}
     \hfill
     \begin{subfigure}[b]{0.32\textwidth}
         \centering
         \includegraphics[width=1\linewidth]{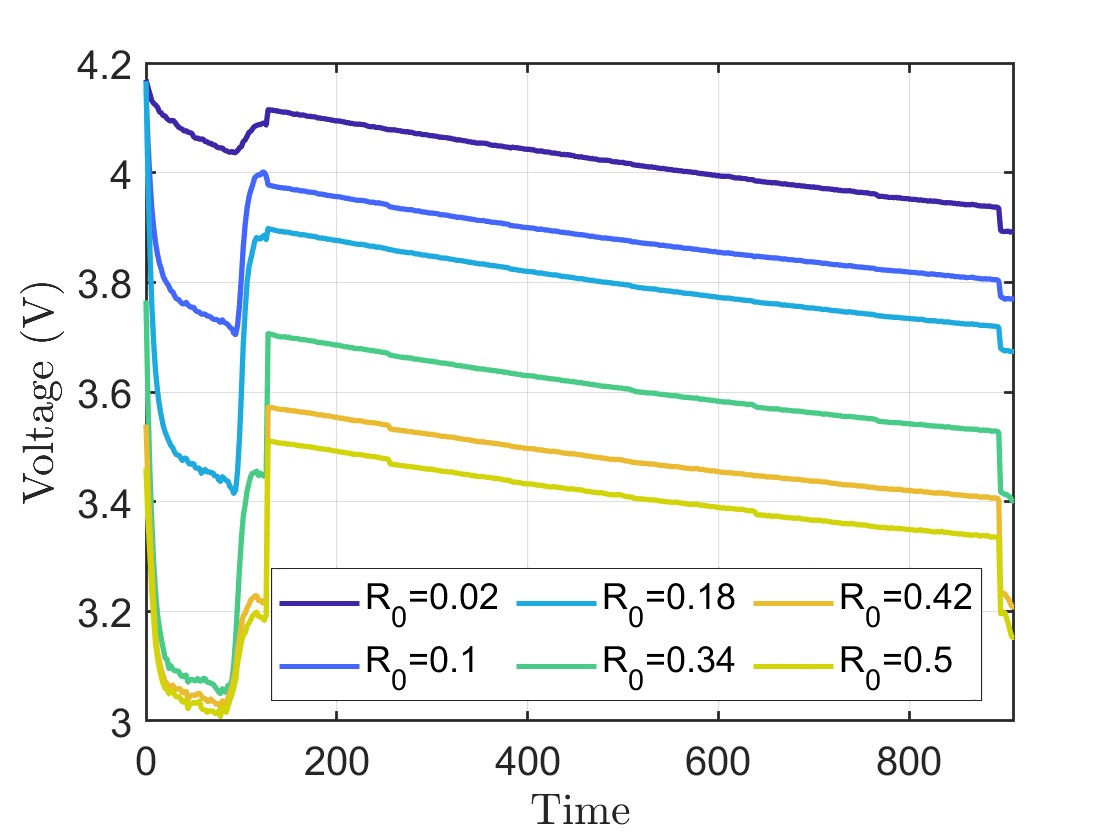}
         \caption{Varying $R_0$ and keeping $q_{max}$ fixed}
           \label{R}
     \end{subfigure}
          \hfill
    \begin{subfigure}[b]{0.32\textwidth}
         \centering
         \includegraphics[width=1\linewidth]{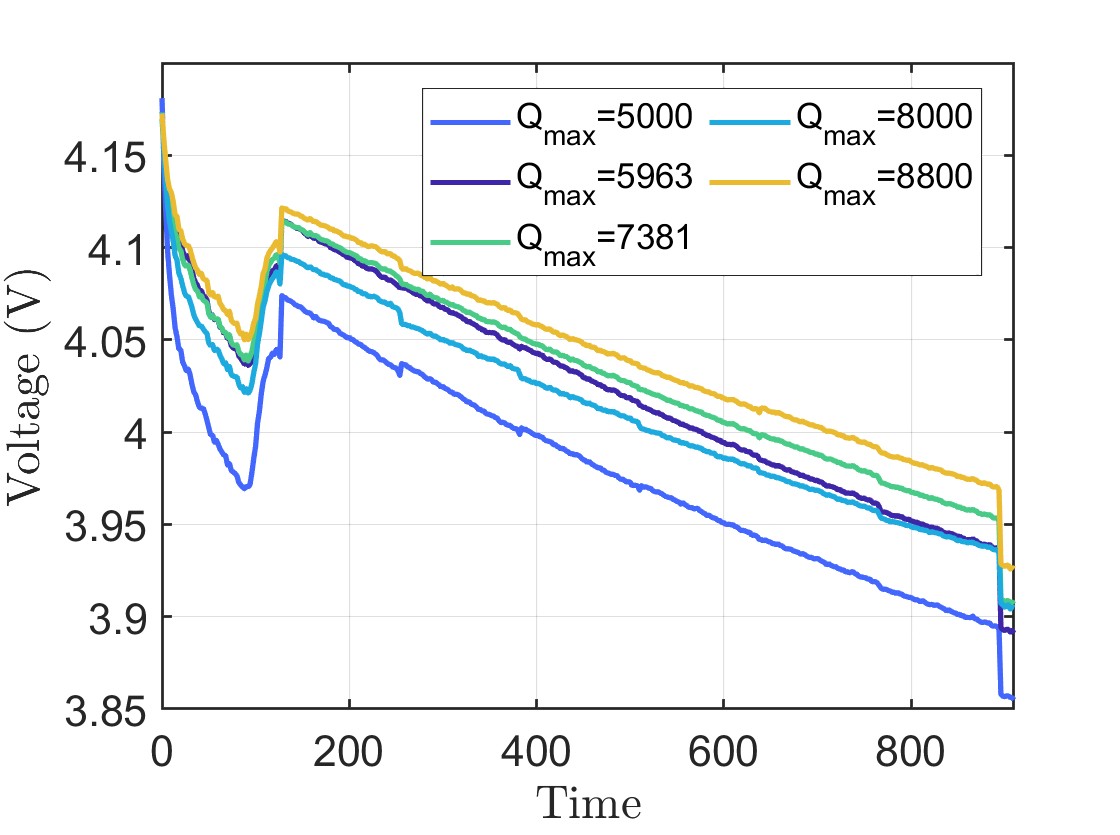}
         \caption{Varying $q_{max}$ and keeping $R_0$ fixed}
           \label{Q}
     \end{subfigure}
              \caption{ Effect of varying degradation parameters on the voltage discharge curves  (with a constant current profile).}
        \label{fig:degradation}
\end{figure}
\begin{figure}[h!]
     \centering
     \begin{subfigure}[b]{0.49\textwidth}
         \centering
         \includegraphics[width=1\linewidth]{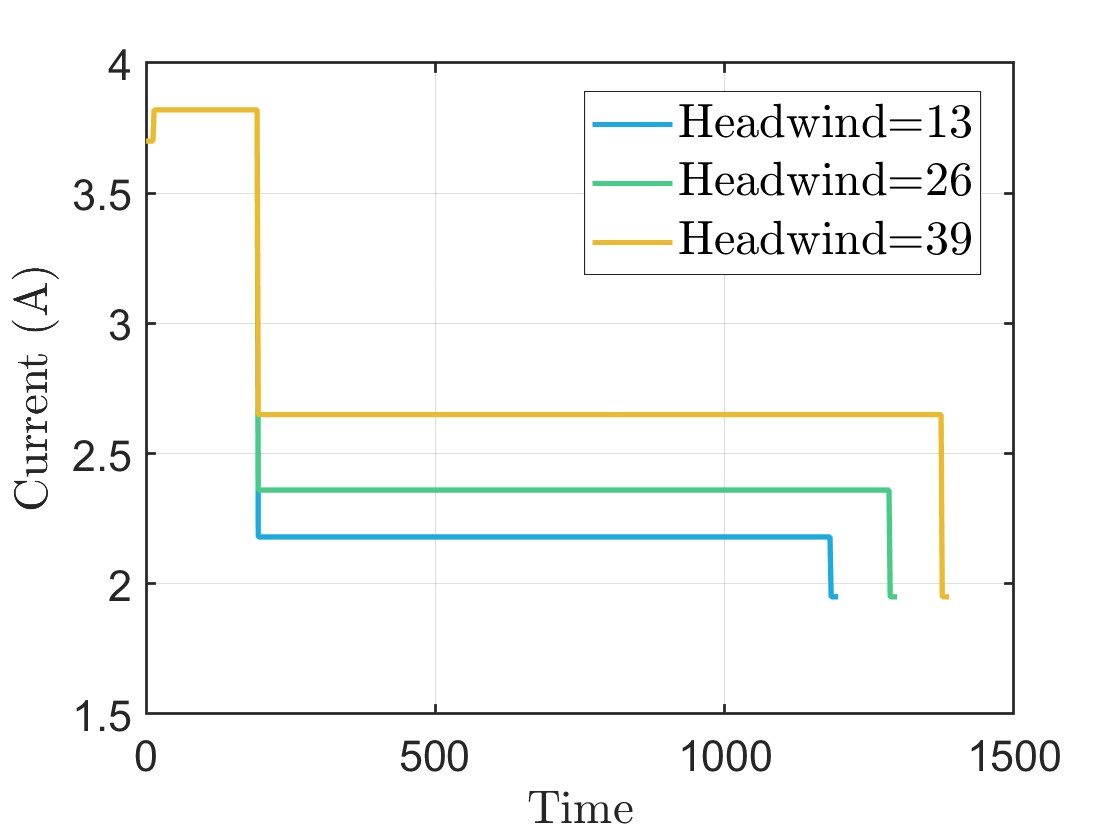}
         \caption{  Current profile for different wind condition}
         \label{current}
     \end{subfigure}
     \hfill
     \begin{subfigure}[b]{0.5\textwidth}
         \centering
         \includegraphics[width=1\linewidth]{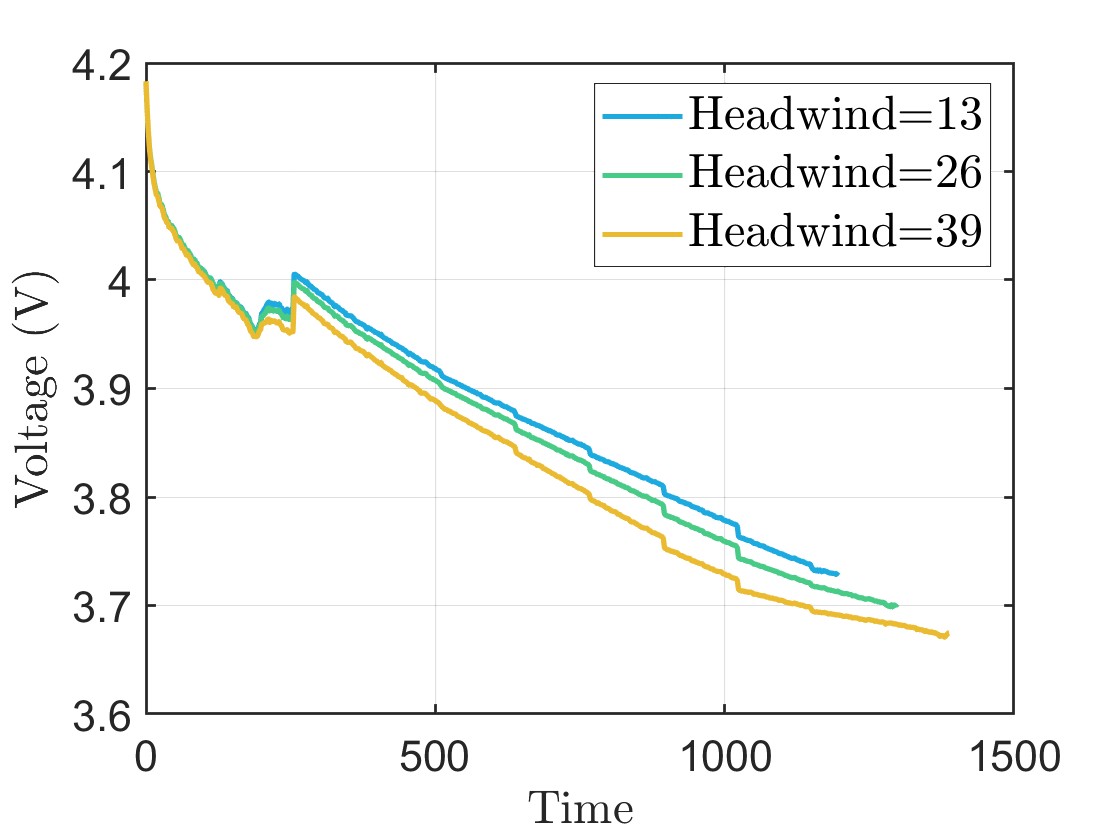}
         \caption{Voltage discharge curves for different wind conditions}
           \label{R}
     \end{subfigure}
              \caption{ Effect of wind condition on the voltage discharge curves and current profile for a specific flight.}
        \label{fig:wind1}
\end{figure}

\begin{figure}[h!]
     \centering
     \begin{subfigure}[b]{0.45\textwidth}
         \centering
         \includegraphics[width=1\linewidth]{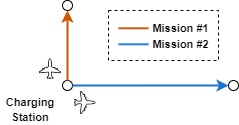}
         \caption{ Map of two single missions}
         \label{map1}
     \end{subfigure}
     \hfill
     \begin{subfigure}[b]{0.45\textwidth}
         \centering
         \includegraphics[width=1\linewidth]{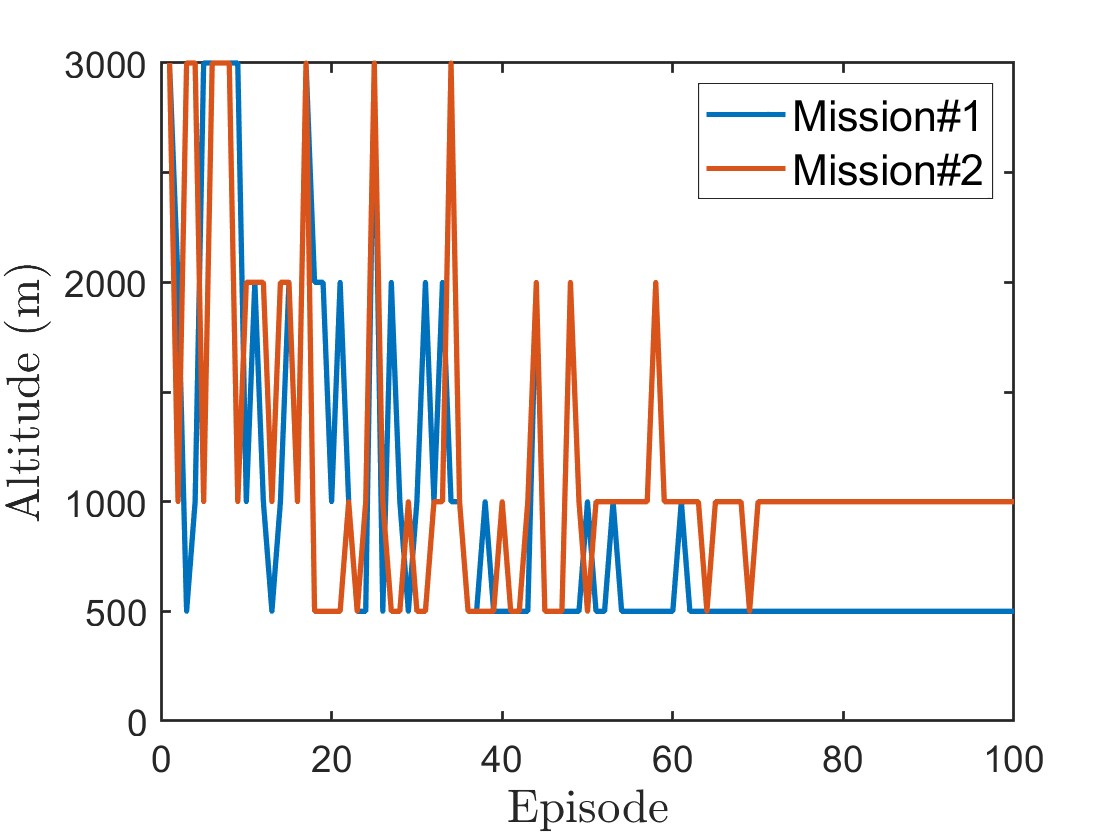}
         \caption{Altitude of aircraft in each mission }
           \label{sim1}
     \end{subfigure}
              \caption{Altitude of aircraft in each mission during each episode using the learning algorithm}
        \label{fig:altitude}
\end{figure}

\subsubsection{Mission with a Single Charging Cycle}
In this scenario, the aircraft is tasked with reaching multiple destinations, each assigned varying priority levels, and returning to the charging station before reaching EOD. Wind conditions may vary for reaching each destination in one mission.
The mission's objective is to reach as many high-priority destinations as possible before returning to the charging station. 
To achieve this, the aircraft must decide on the destinations to reach, the order in which to reach them, and the altitude at which to fly to each destination.
Figure \ref{mission2} provides an illustrative example of such a mission, depicting a scenario where the aircraft faces eight destinations. In this case, the aircraft strategically selects destinations \#1, 2, 3, and 6 based on their priority levels, the current battery health state, and wind conditions. In the following, we will discuss interesting observations and results that we achieved in this mission.  

\begin{figure}[h!]
     \centering
         \centering
         \includegraphics[width=0.6\linewidth]{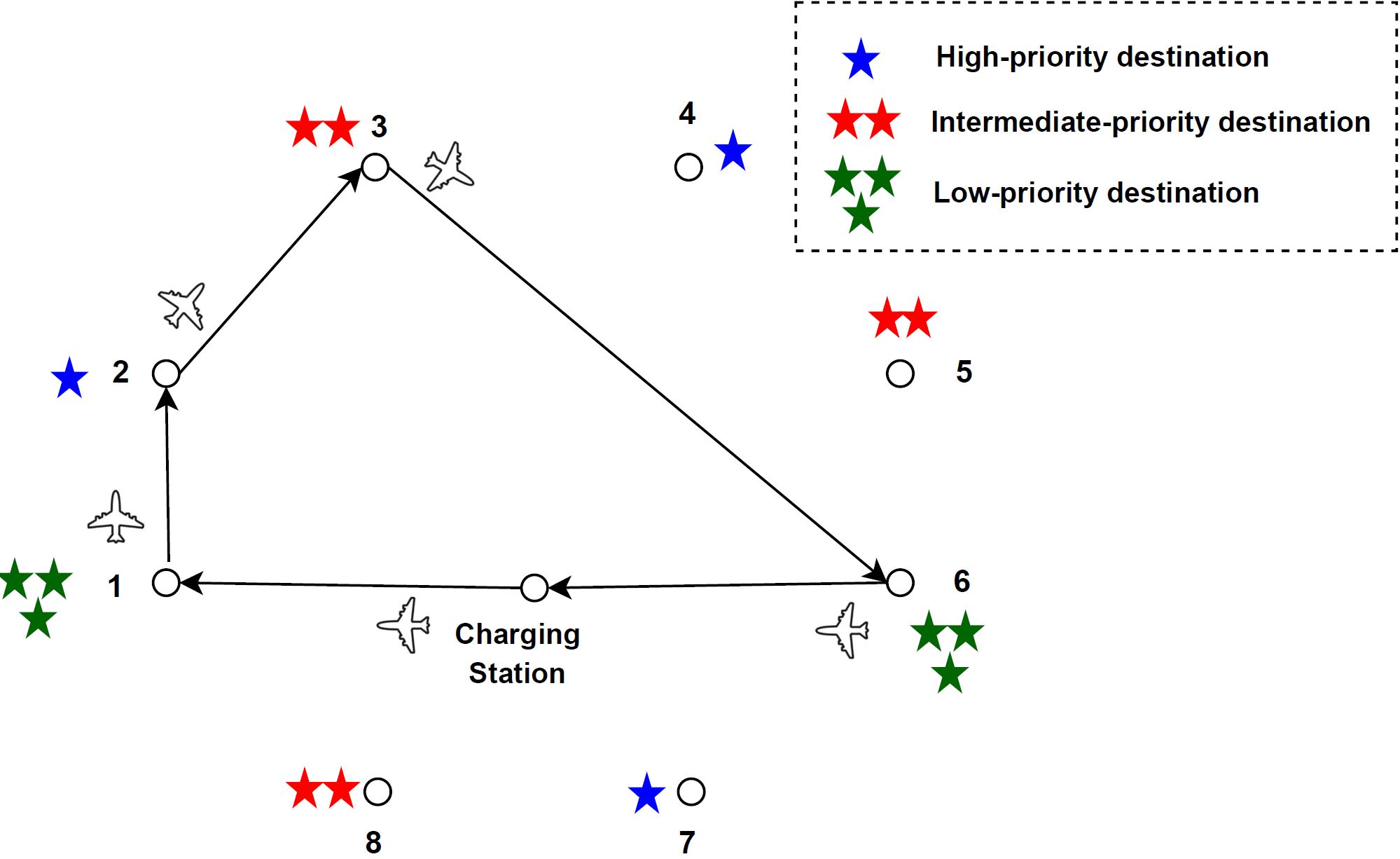}
         \caption{  A mission with a single charge of battery. The number of stars next to each destination shows the priority of that destination.}
         \label{mission2}
\end{figure}
Figure \ref{fig:mis2exp}  provides a more detailed visualization of this mission. As shown in Figure \ref{altitude=3000}, the aircraft can only reach a single destination if it maintains a cruise altitude of 3000 m. However, during the flight path to the second destination, the voltage of the aircraft's battery falls below the EOD threshold. On the other hand, by adjusting the cruise altitude to 1000 m, the aircraft becomes capable of reaching an additional destination, allowing it to visit two destinations before reaching the EOD limit (Figure \ref{altitude=1000}). It is important to note that in both of these 
missions, the headwind is set to 39 kts.
\begin{figure}[h!]
     \centering
     \begin{subfigure}[b]{0.32\textwidth}
         \centering
         \includegraphics[width=1\linewidth]{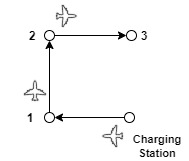}
         \caption{  Map of an example mission}
         \label{map_m2}
     \end{subfigure}
     \hfill            
    \begin{subfigure}[b]{0.32\textwidth}
         \centering
         \includegraphics[width=1\linewidth]{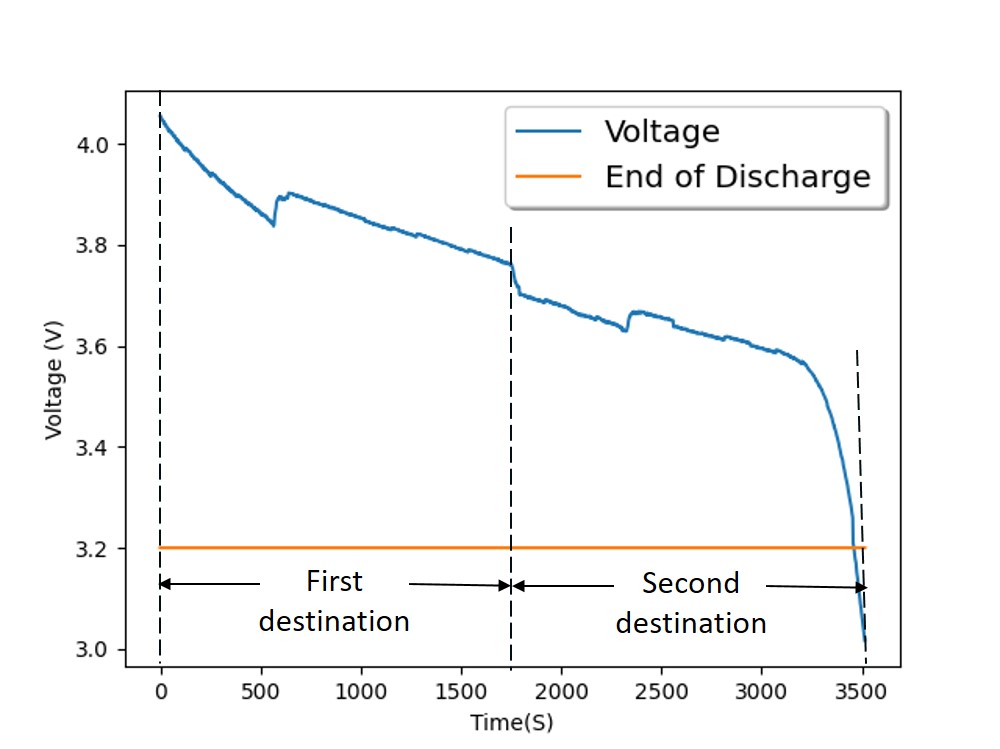}
         \caption{Voltage discharge curve for cruise altitude=3000 m}
           \label{altitude=3000}
     \end{subfigure}
       \hfill
     \begin{subfigure}[b]{0.32\textwidth}
         \centering
         \includegraphics[width=1\linewidth]{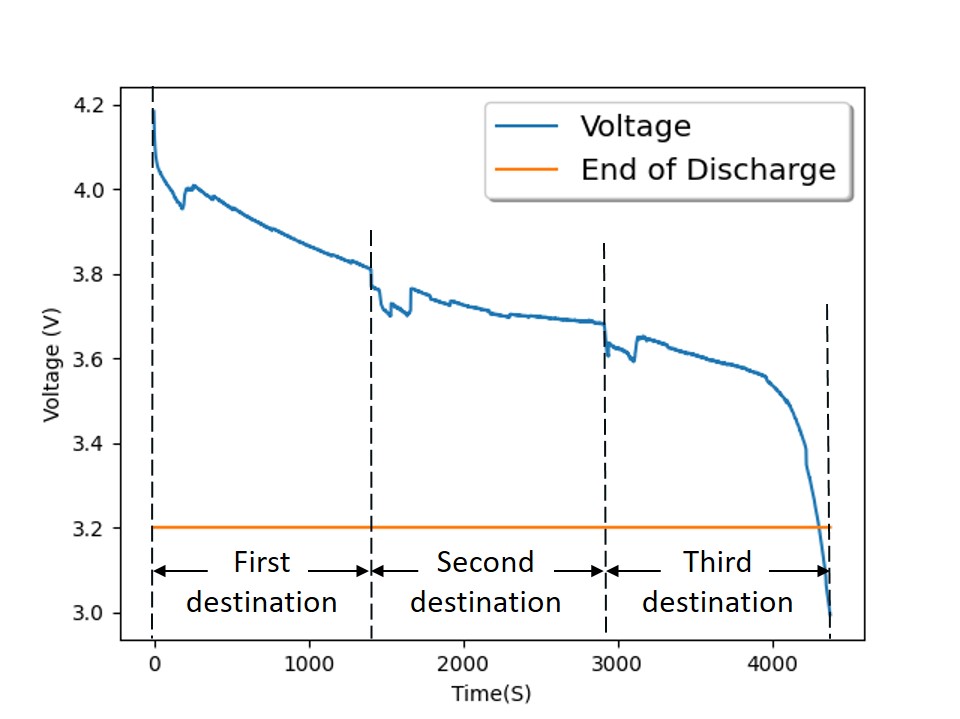}
         \caption{Voltage discharge curve for cruise altitude=1000 m }
           \label{altitude=1000}
     \end{subfigure}
              \caption{ Impact of flight altitude on voltage discharge curve and the number of destinations that the aircraft reached before reaching EOD}
        \label{fig:mis2exp}
\end{figure}
\begin{figure}[h!]
     \centering
     \begin{subfigure}[b]{0.49\textwidth}
         \centering
         \includegraphics[width=1\linewidth]{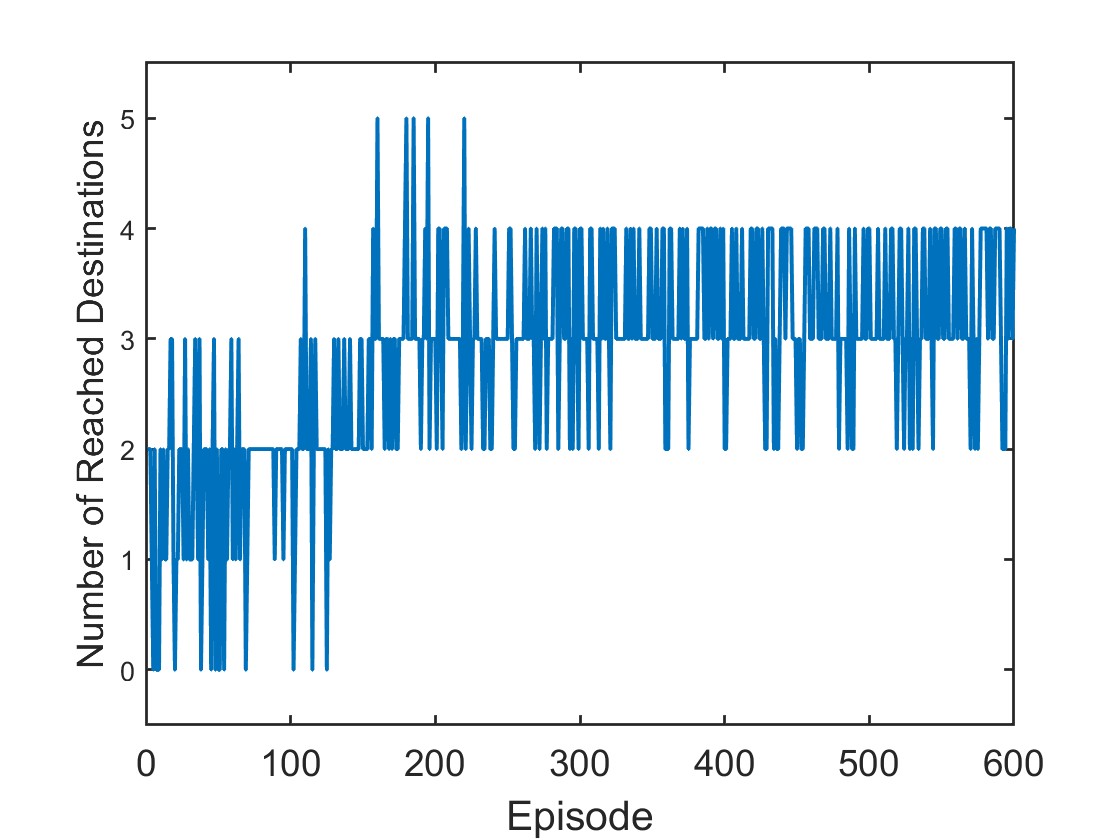}
         \caption{Number of reached destinations}
         \label{number2}
     \end{subfigure}
     \hfill
     \begin{subfigure}[b]{0.5\textwidth}
         \centering
         \includegraphics[width=1\linewidth]{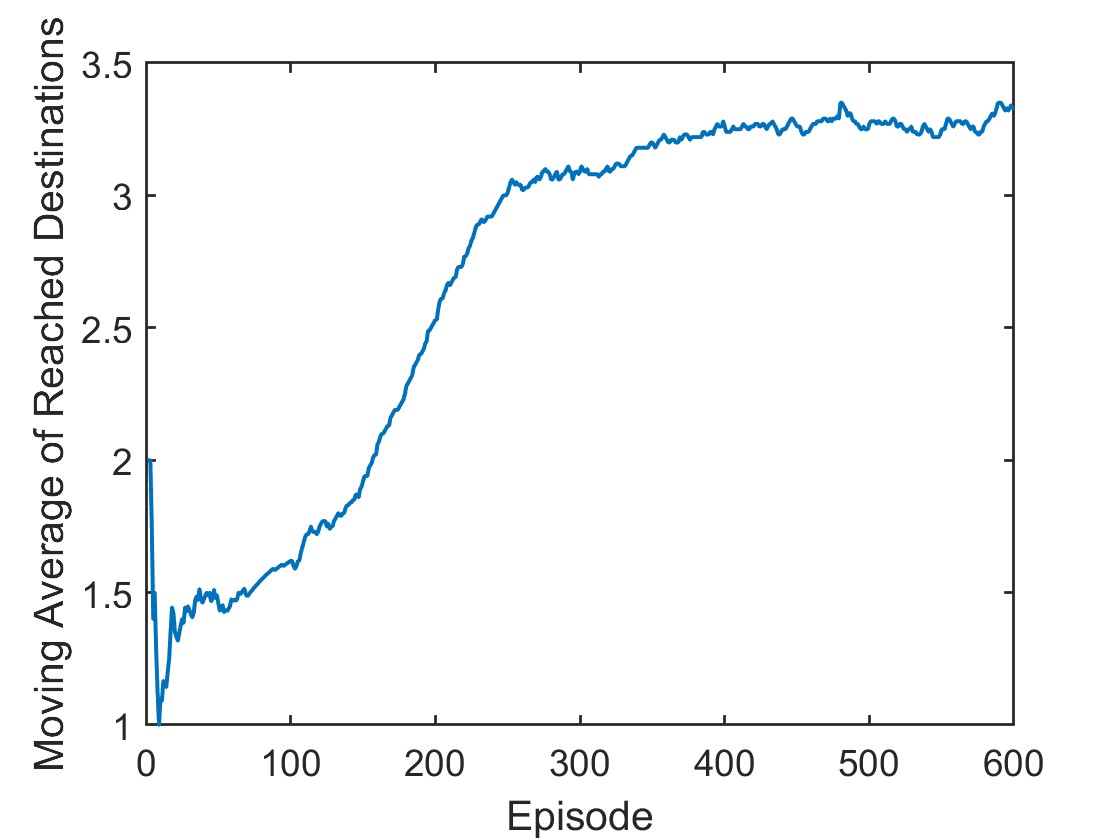}
         \caption{moving average of the number of reached destinations}
           \label{number2MA}
     \end{subfigure}
              \caption{Number of reached destinations and moving average of the number of reached destinations in each episode using a learning algorithm.}
        \label{fig:mission21}
\end{figure}

To illustrate the learning progress of the aircraft using the proposed DRL method in this scenario, we plot the number of destinations the aircraft reaches before returning to the charging station or reaching the EOD in each learning episode. This is depicted in Figure \ref{number2}. 
The results demonstrate that the aircraft gradually improves its performance and achieves more destinations over subsequent episodes. 
However, it is important to note that due to random sampling of the battery's health and wind conditions for each flight, the number of destinations that the aircraft reaches may vary even after the aircraft learns its optimal strategy. 
To provide a more comprehensive representation of the learning process, we compute the moving average of the number of destinations reached for each learning episode, as depicted in Figure \ref{number2MA}. It is noteworthy that while Figure \ref{number2} indicates instances where the aircraft reaches five destinations, these cases result in the battery voltage reaching EOD before reaching the charging station. As a result, the aircraft incurs a substantial negative reward and learns to avoid such conditions in subsequent episodes.
\subsubsection{Mission with Multiple-Charging Cycles}
In the third scenario, the aircraft's mission is to reach all the predefined destinations with a minimum number of charging cycles.
 To accomplish this, the aircraft must make decisions regarding the order of destinations to reach, the timing of returning to the charging station, and the altitude at which to fly to reach each destination. Figure~\ref{mission3} shows an example of this kind of mission when the aircraft faced a mission with eight destinations. \\
\begin{figure}[h!]
         \centering
         \includegraphics[width=0.6\linewidth]{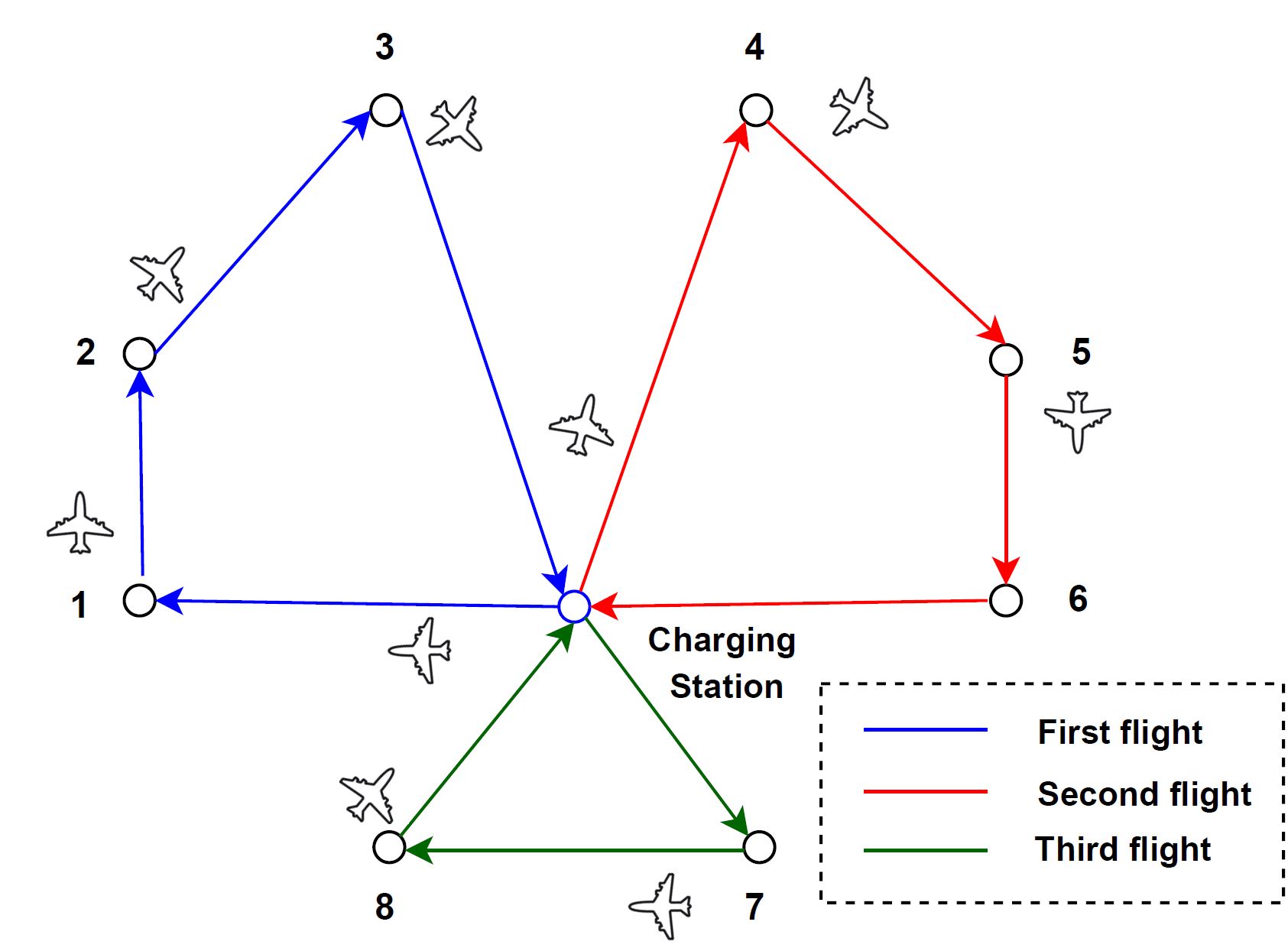}
         \caption{A mission with a multi-charge of battery. Each color represents the path that the aircraft flights in one charging cycle.}
           \label{mission3}
        \label{fig:wind}
\end{figure}
To illustrate the learning behavior of the aircraft through the proposed DRL method in this scenario, we plot the number of destinations reached by the aircraft before reaching the EOD (Figure \ref{distin}) and the number of charging cycles throughout each learning episode in this mission (Figure \ref{cycle}). 
As illustrated in these figures, in the initial episodes, the aircraft's battery voltage drops below the EOD threshold before reaching all destinations. However, with training time, the aircraft learns the optimal order of destinations, the timing for returning to the charging station, and the optimal flight altitude to reach all destinations without encountering EOD issues. Given the significant penalty associated with falling below the EOD threshold, the aircraft prioritizes reaching all destinations without reaching EOD, even if it requires more charging cycles. Subsequently, the aircraft shifts its focus towards minimizing the number of charging cycles while still reaching all destinations. This dynamic is reflected in Figure \ref{cycle}, where we observe an initial increase followed by a decrease in the number of charging cycles as the learning progresses. It is noticeable that to prevent the aircraft from learning to immediately return to the charging station after reaching each destination, it is crucial to finely tune the parameters of the reward functions, denoted as $\alpha$, $\beta$, and $\gamma$ in Equation \ref{eq:reward}. By carefully adjusting these parameters, we can effectively shape the learning process and optimize the aircraft's decision-making behavior. In the following, we investigate the effect of these parameters on the learning process.

\begin{figure}[h!]
     \centering
     \begin{subfigure}[b]{0.49\textwidth}
         \centering
         \includegraphics[width=1\linewidth]{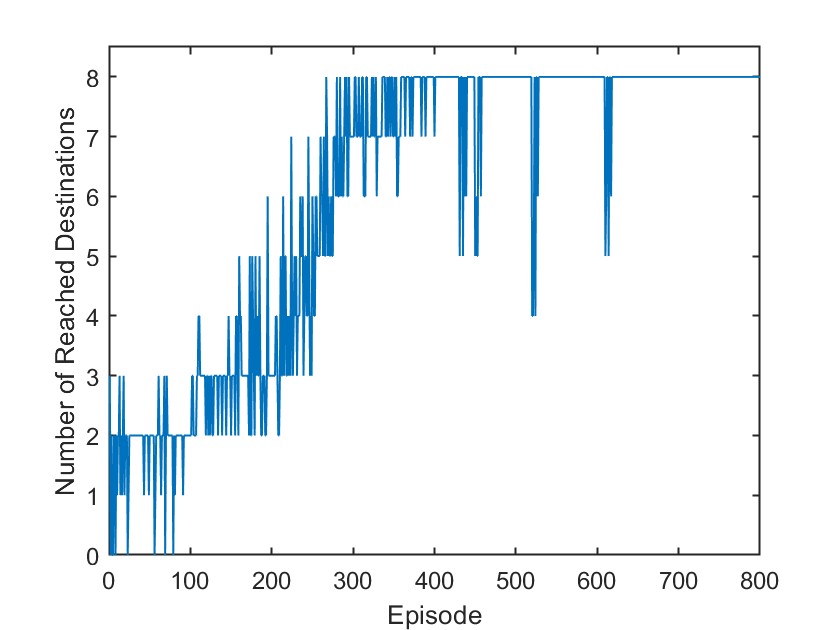}
         \caption{Number of reached destinations in each episode}
         \label{distin}
     \end{subfigure}
     \hfill
     \begin{subfigure}[b]{0.5\textwidth}
         \centering
         \includegraphics[width=1\linewidth]{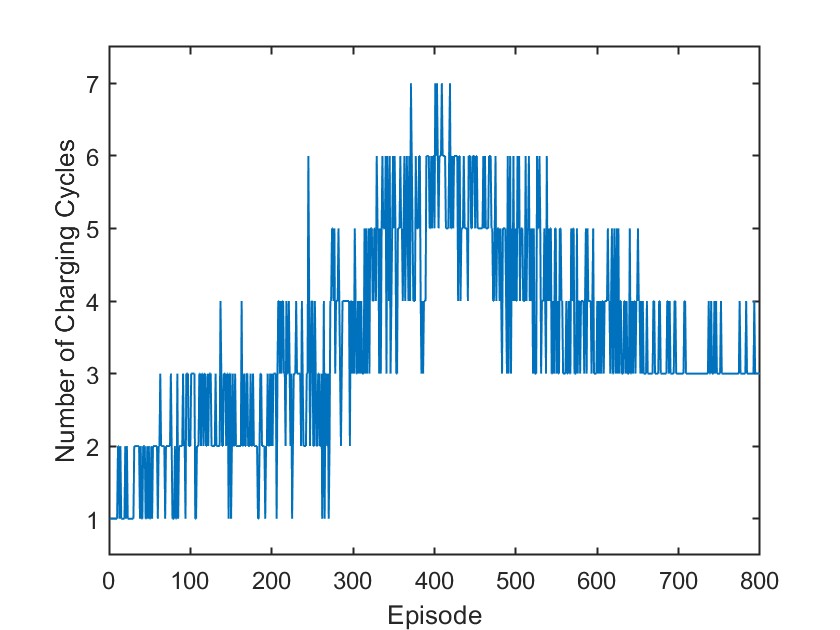}
         \caption{Number of charging cycles in each episode}
           \label{cycle}
     \end{subfigure}
              \caption{ Number of reached destinations and charging cycles in each episode using learning algorithm}
        \label{fig:mission3}
\end{figure}
As outlined in Section \ref{sec:Reinforcement learning}, the aircraft faces a substantial penalty should its battery voltage fall below the EOD threshold level before reaching the destination. The impact of this penalty, denoted by $\gamma$ in equation \ref{eq:reward}, on both the number of reached destinations and the number of charging cycles is visually demonstrated in Figures \ref{comgamma1} and \ref{comgamma2}. Note that throughout these figures, $\beta$ which represents the penalty imposed when the aircraft reaches a charging station, and $\alpha$  which signifies the reward for reaching each destination, remain fixed.
Figure \ref{comgamma1} reveals that as $\gamma$ increases, the aircraft becomes more adept at reaching all destinations in fewer learning episodes without breaching the EOD threshold. As shown in Figure \ref{comgamma2}, in the early episodes, the aircraft learns to reach all destinations by maximizing the number of charging cycles. This strategic approach ensures the avoidance of penalties associated with the battery voltage dropping below the EOD level. Subsequently, the aircraft gradually refines its strategy, learning to reach all destinations with the minimum number of charging cycles.
Moreover, Figure \ref{comgamma2} shows that when increasing $\gamma$, the aircraft's learning process to reach all destinations with the minimum number of charging cycles requires more time. Furthermore,  with higher $\gamma$ values, the aircraft tends to adopt a more conservative approach, opting for additional charging cycles to reach all destinations. This conservative strategy ensures that irrespective of the battery health state or wind conditions, the voltage will not fall below the EOD level.

Similarly, the aircraft faces a penalty upon reaching a charging station, albeit less severe than the penalty incurred when the battery voltage falls below the EOD level. This penalty structure serves to incentivize the aircraft to efficiently reach multiple destinations within a charging cycle, thereby optimizing its overall performance.
The impact of this penalty, denoted by $\beta$ in equation \ref{eq:reward}, on both the number of reached destinations and the number of charging cycles is visualized in Figures \ref{combeta1} and \ref{combeta2}. It is crucial to emphasize that throughout these figures, both $\gamma$ and $\alpha$ remain constant.
Figure \ref{combeta2} demonstrates that as the penalty parameter $\beta$ increases, the aircraft progressively refines its strategy, becoming more proficient at reaching all destinations with the minimum number of charging cycles. Notably, when $\beta=0$, the aircraft exhibits minimal learning regarding the optimization of charging cycles. In this case, since the aircraft returns to the charging station after reaching each destination, the battery voltage seldom drops below the EOD. Consequently, as depicted in Figure \ref{combeta1}, after the aircraft learns its optimal strategy, the aircraft consistently reaches all destinations.
\begin{figure}[h!]
         \centering
         \includegraphics[width=1\linewidth]{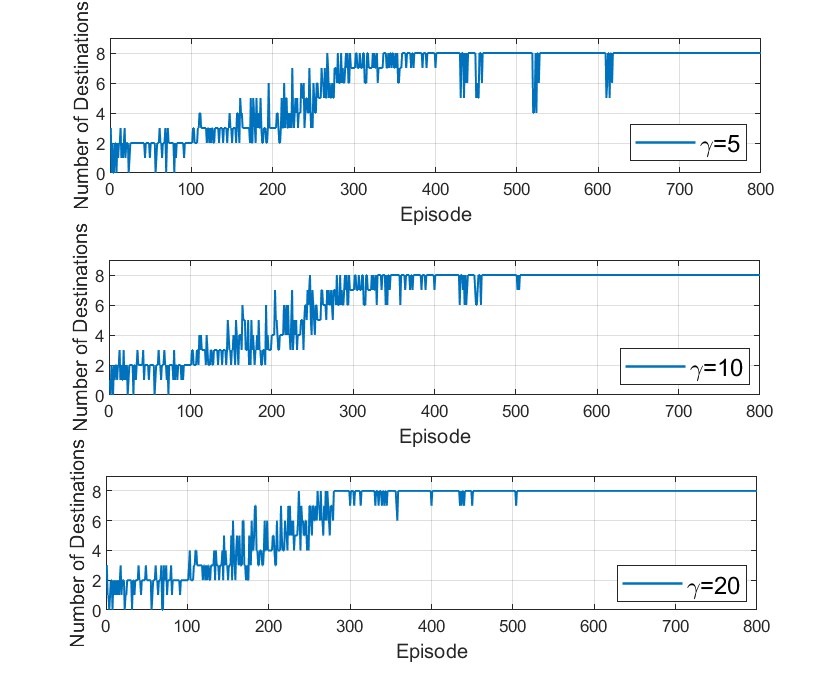}
         \caption{Number of reached destinations in each episode for different values of $\gamma$ in reward function while both $\alpha=1$ and $\beta=1$ remain fixed.}
           \label{comgamma1}
\end{figure}
\begin{figure}[h!]
         \centering
         \includegraphics[width=1\linewidth]{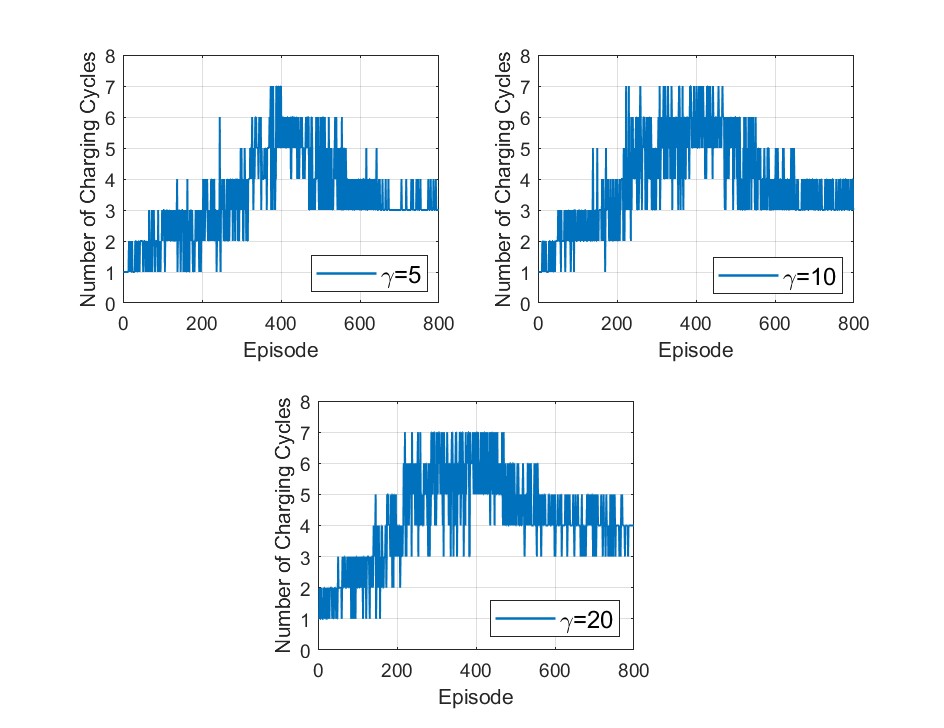}
         \caption{Number of charging cycles in each episode for different values of $\gamma$ in reward function while both $\alpha=1$ and $\beta=1$ remain fixed. }
           \label{comgamma2}
        \label{fig:wind}
\end{figure}
\begin{figure}[h!]
         \centering
         \includegraphics[width=1\linewidth]{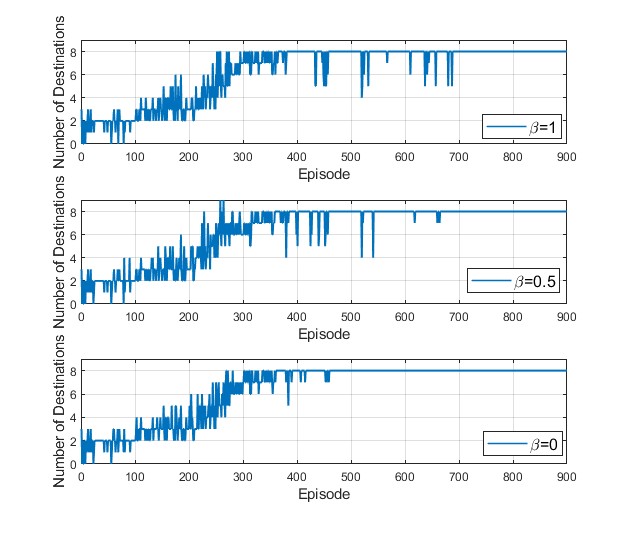}
         \caption{Number of reached destinations in each episode for different values of $\beta$ in reward function while both $\alpha=1$ and $\gamma=5$ remain fixed. }
           \label{combeta1}
\end{figure}
\begin{figure}[h!]
         \centering
         \includegraphics[width=1\linewidth]{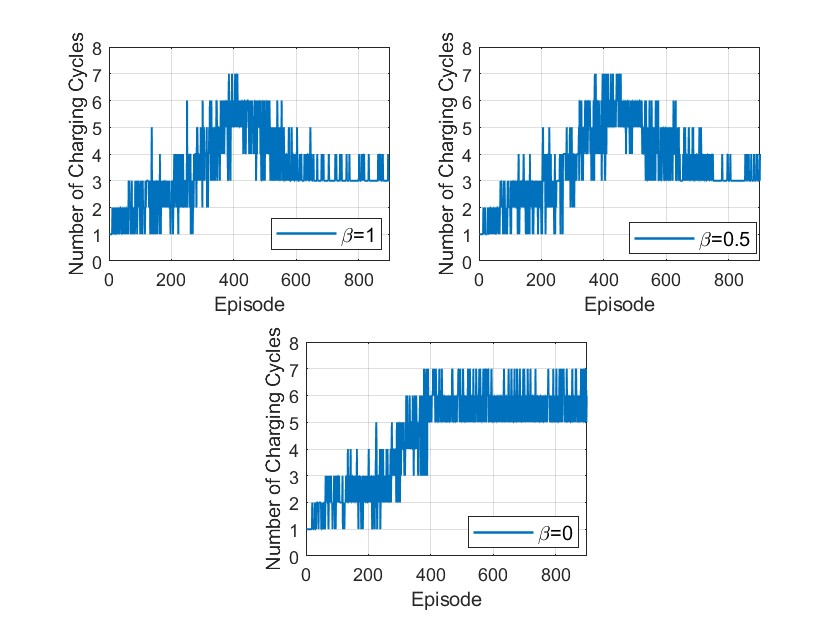}
         \caption{Number of charging cycles in each episode for different values of $\beta$ in reward function while both $\alpha=1$ and $\gamma=5$ remain fixed. }
           \label{combeta2}
        \label{fig:wind}
\end{figure}
\section{Conclusion}
\label{sec:Conclusion}
In summary, realizing the transformative potential of Urban Air Mobility depends on the successful integration of battery-powered aircraft into urban transportation systems. The widespread adoption of UAM relies on overcoming challenges associated with battery degradation and unforeseen events, both critical elements in mission planning and health-aware real-time operational control.

Our research has addressed the complex task of combining mission planning and health-aware real-time operational control, with a particular focus on integrating battery health and degradation state into optimal decision-making. The proposed DRL algorithm represents a proactive strategy, prescribing operational parameters that extend the battery's discharge cycle based on its current health status, while simultaneously maximizing mission efficiency. 
In numerous scenarios simulated with a NASA multirotor aircraft model, the proposed deep reinforcement learning framework demonstrates its adaptability and versatility.

The proposed algorithm's strength lies in its ability to adapt to batteries in different states of degradation without any prior knowledge of the true degradation state and effectively handle uncertainties, a critical feature in the unpredictable UAM environment where factors like adverse weather conditions can disrupt operations. While predicting the end of discharge can already be very advantageous,  our proposed algorithm goes a step further by incorporating information on the health condition into its decision-making process. The algorithm optimizes operations in a health-aware manner, ensuring near-optimal performance across diverse operational scenarios and efficient aircraft operation. Beyond individual aircraft optimization, our approach has the potential to enhance overall system throughput by optimizing fleet operations. As UAM continues to evolve, the adaptability and robustness of our algorithm position it as a valuable tool for ensuring the reliable and efficient integration of battery-powered aircraft into urban transportation networks.
The proposed prescriptive algorithm is characterized by its low computational cost and fast processing time, enabling seamless integration into air traffic control systems and facilitating prompt and efficient decision-making.


Our work can be extended in multiple ways. Firstly, the health-aware control strategy devised in this study provides a versatile framework applicable beyond Urban Air Mobility. Future investigations could explore the adaptability of the proposed framework in electric vehicles, portable electronic devices, and renewable energy systems. Secondly, the integration of the proposed algorithm into air traffic management systems has the potential to alleviate airspace congestion and improve traditional air traffic. Furthermore, optimal charging station placement in urban areas could be a subject of exploration utilizing optimization algorithms. Lastly, there is potential in relaxing some of the limiting assumptions considered in this research, such as the finite set of wind conditions, the finite set of altitudes that the aircraft can fly to reach each destination, or the assumption of constant wind conditions throughout the flight to reach each destination.

\bibliography{bib_items,sample}
\bibliographystyle{ieeetr}
\appendix 
\label{Appendix}

\section{Simulation Setup}
\label{apB}
Table \ref{profile_power} provides an overview of the power required (in kW) during various phases of the 30 nm flight. These phases encompass vertical takeoff, climbing, cruise, approach, and vertical landing, with corresponding phase duration detailed in Table \ref{profile_time}.
\begin{table*}[th]
    \begin{center}
    \footnotesize
        \caption{Power required (kW) in different phases of flight of a simulated 30 nm flight of aircraft in 13 kts tailwind }
        \footnotesize
        \label{profile_power}
        \begin{tabular}[t]{c|cccccc}
        \toprule
    \multicolumn{1}{}{} &\multicolumn{6}{|c}{Power Required (kW) } \\
        \hline
        {Cruise Altitude (m)} & Vertical Takeoff & Climb &  Cruise &Descent (45 Deg) &Approach (8 Deg) &Vertical Landing\\
        \hline
        500  &  264.94 & 273.53&138.39  &Negligible&Negligible&139.45
\\
        1000  &  264.94 & 272.66&141.72 &Negligible&Negligible&139.45
\\
        2000 &  264.94 & 273.03&145.68 &Negligible&Negligible&139.45
 \\
        3000 &  264.94& 273.13&155.44 &Negligible&Negligible&139.45
 \\
        \hline
        \end{tabular}
    \end{center}
\end{table*}
\begin{table*}[th]
\footnotesize
    \begin{center}
        \caption{Duration (in seconds) of the different phases of flight of a simulated 30 nm flight of aircraft in 13 kts tailwind }
        \label{profile_time}
        \begin{tabular}[t]{c|cccccc}
        \toprule
    \multicolumn{1}{}{} &\multicolumn{6}{|c}{Flight Duration (sec) } \\
        \hline
        {Cruise Altitude (m)} & Vertical Takeoff & Climb &  Cruise &Descent (45 Deg) &Approach (8 Deg) &Vertical Landing\\
          \hline
        500  &  15.51 & 85
& 1167
 &0&115
&15.5

\\
        1000  &  15.51
 & 180
&1143
 &11.4
&115
&15.5
\\
        2000 &  15.51
 & 369
&1120
 &32.6
&115
&15.5
 \\
        3000 &  15.51
&558
& 1056
&52.1
&115
&15.5
 \\
         \hline
        \end{tabular}
    \end{center}
\end{table*}
In this study, the wind is assumed to be uniform in each flight, i.e., constant magnitude and direction.  The study assesses how wind magnitude uncertainties affect both power requirements and the time of the flight's cruise phase. Tables \ref{wind500}, \ref{wind1000}, \ref{wind2000}, and \ref{wind3000} provide a detailed analysis of how wind affects power requirements and cruise duration at various cruise altitudes.
\begin{table*}[th]
\footnotesize
    \begin{center}
        \caption{Power required (kW) and duration of flight (sec) in cruise phase when the cruise altitude is 500m }
        \label{wind500}
        \begin{tabular}[t]{cccc}
        \toprule
        \hline
        Wind Magnitude (kts) & Cruise Airspeed & Power Required (kW)& Flight Duration (sec)\\
        \hline
        -39  &  38 m/s (73.9 kts) & 130.57&959 \\
        -26  &  39 m/s (75.8 kts) & 132.92&1063 \\
        -13 &  41 m/s (79.7 kts) & 138.39&1167 \\
        13 &  45 m/s (87.5 kts)& 152.18&1453 \\
        26 &  48 m/s (93.3 kts)& 165.22&1608 \\
        39 &  52 m/s (101.1 kts)& 186.36&1743 \\
        \hline
        \end{tabular}
    \end{center}
\end{table*}
\begin{table*}[th]
\footnotesize
    \begin{center}
        \caption{Power required (kW) and duration of flight (sec) in cruise phase when the cruise altitude is 1000m }
        \label{wind1000}
        \begin{tabular}[t]{cccc}
        \toprule
        \hline
        Wind Magnitude (kts) & Cruise Airspeed & Power Required (kW)& Flight Duration (sec)\\
        \hline
        -39  &  39 m/s (75.8 kts) & 133.78&943 \\
        -26  & 40 m/s (77.7 kts) & 136.18&1043 \\
        -13 &  42 m/s (81.6 kts) & 141.72&1143 \\
        13 &  46 m/s (89.4 kts)& 155.59&1416 \\
        26 &  49 m/s (95.2 kts)& 168.61&1563 \\
        39 &  53 m/s (103.1 kts)& 189.71&1690 \\
        \hline
        \end{tabular}
    \end{center}
\end{table*}
\begin{table*}[th]
\footnotesize
    \begin{center}
        \caption{Power required (kW) and duration of flight (sec) in cruise phase when the cruise altitude is 2000m }
        \label{wind2000}
        \begin{tabular}[t]{cccc}
        \toprule
        \hline
        Wind Magnitude (kts) & Cruise Airspeed & Power Required (kW)& Flight Duration (sec)\\
        \hline
        -39  &  41 m/s (79.7 kts) & 140.75&911 \\
        -26  & 42 m/s (81.6 kts) & 143.01&1005 \\
        -13 & 43 m/s (93.3 kts) & 148.68&1120 \\
        13 &  48 m/s (93.3 kts)& 162.21&1347 \\
        26 &  51 m/s (99.1 kts)& 174.99&1479 \\
        39 & 55 m/s (106.9 kts)& 195.3&1594 \\
        \hline
        \end{tabular}
    \end{center}
\end{table*}
\begin{table*}[th]
\footnotesize
    \begin{center}
        \caption{Power required (kW) and duration of flight (sec) in cruise phase when the cruise altitude is 3000m }
        \label{wind3000}
        \begin{tabular}[t]{cccc}
        \toprule
        \hline
        Wind Magnitude (kts) & Cruise Airspeed & Power Required (kW)& Flight Duration (sec)\\
        \hline
        -39  &  43 m/s (83.6 kts) & 147.82&882 \\
        -26  & 44 m/s (85.5 kts) & 150.07&970 \\
        -13 & 46 m/s (89.4 kts) & 155.44&1056 \\
        13 & 50 m/s (97.1 kts)& 172.61&1256 \\
        26 &  53 m/s (103.0 kts)& 181.03&1405 \\
        39 & 57 m/s (110.8 kts)& 200.77&1507\\
        \hline
        \end{tabular}
    \end{center}
\end{table*}
\end{document}